# Violent music vs violence *and* music:
# Drill rap and violent crime in London

Bennett Kleinberg[1,2] and Paul McFarlane[1,3]

[1] Department of Security and Crime Science, University College London
[2] Dawes Centre for Future Crime, University College London
[3] Jill Dando Institute for Global City Policing, University College London[1]

**Abstract**   The current policy of removing drill music videos from social media platforms such as YouTube remains controversial because it risks conflating the co-occurrence of drill rap and violence with a causal chain of the two. Empirically, we revisit the question of whether there is evidence to support the conjecture that drill music and gang violence are linked. We provide new empirical insights suggesting that: i) drill music lyrics have not become more negative over time if anything they have become more positive; ii) individual drill artists have similar sentiment trajectories to other artists in the drill genre, and iii) there is no meaningful relationship between drill music and 'real-life' violence when compared to three kinds of police-recorded violent crime data in London. We suggest ideas for new work that can help build a much-needed evidence base around the problem.

**INTRODUCTION**
The role of drill music – a subgenre of aggressive rap music (Mardean, 2018) – in inciting gang-related violence remains controversial. The current policy towards the drill-gang violence nexus is still somewhat distant from understanding the complexity of drill music language and how drill music videos are consumed by a wide range of social media users. To build the evidence base for more appropriate policy interventions by the police and social media companies, our previous work looked at dynamic sentiment trajectories (i.e. the use of positive language throughout a document) to provide the first empirical glimpse at drill music language from a computational angle (Kleinberg and McFarlane, 2019).

---

[1] Contact: bennett.kleinberg@ucl.ac.uk and p.mcfarlane@ucl.ac.uk

Our findings recommended it would be prudent for policymakers to stay close to the evidence for two key reasons. Firstly, videos with an overly negative language sentiment did not attract more views or create more engagement on YouTube and, secondly, videos that contained a more positive sentiment received twice as many views on YouTube. Significantly, social media users engaged with these positive videos more than twice as much as songs in the negative groups (Kleinberg and McFarlane, 2019; McFarlane and Kleinberg, 2020).

The potential problem around drill music is not unique to the UK. The Netherlands has recently experienced an increase in the production of and media attention to youth "gangs" involved in the production of drill rap. Similar to the UK narrative, incidents of violence are connected to drill music. Thus, on the one hand, we have a youth culture phenomenon that has risen to prominence in Chicago (Mardean, 2018), London (Dearden, 2018) and the Dutch cities of Amsterdam (Borst and Tourkov, 2019) and Rotterdam (de Haan, 2020), and is seen as the voice of youths in urban districts. On the other hand, however, there are incidents of severe violence in these areas that some policymakers connect to the drill music genre. Author Will Ashon, when asked about the role of aggressive rap music in violence, stated in an interview with the Dutch *Volkskrant* (translated) that "[r]eal physical violence is always problematic. Making rap music about it is a different story. In the UK, there is a lot of commotion about drill music […]. As always, the music is seen as the culprit while drill rap and gang violence are symptoms of the same underlying problems: poverty, lack of opportunity, poor housing, lack of education, etc." (Kraak, 2018). Whether or not that nexus exists is an empirical question, and this work aims to further add to the evidence base around the controversial questions that have arisen from the two co-occurring trends.

*The current study*
In this paper, we address the limitations of our previous work by reducing the frequency of out of vocabulary terms (OOV) and expand our study of drill lyrics to provide new empirical insights both on the genre-level as well as on the artist-level. We find suggestive evidence in support of two mechanisms: drill music lyrics have not become more negative over time (if anything they have become more positive), and individual drill artists have similar sentiment trajectories to other artists in the drill genre. In a preliminary analysis, we do not find evidence of a relationship between drill music and 'real-life' violence when compared to three violent crime trajectories in London.

**DATA**

*Drill lyrics data*
We utilized the data from previous work on drill music (Kleinberg and McFarlane, 2019). Specifically, the dataset consists of the lyrics of 105 different drill music artists affiliated with some gang activity on London. The total corpus size was 378,822 tokens with 549 different songs ($M_{length}$ = 688.77, $SD$ = 235.21).[2] For details on the data retrieval process, see (Kleinberg and McFarlane, 2019). The temporal range of the data (i.e. the publishing date of the video on YouTube) was Dec. 2013 to Nov. 2018.

---

[2] Note that in contrast to the original paper, we removed the oldest video from Nov. 2007 not to add periods of in-activity to the analysis.

*Crime data*

Our study used data relating to three violent crime types count trajectories in London to determine whether there are any relationships with the temporal progression of the drill music corpus. These crime counts trajectories included homicide, robbery of personal property and violence with injury from December 2013 to November 2018.[3] These data were downloaded from the publicly available Metropolitan Police crime data dashboard (see: https://www.met.police.uk/sd/stats-and-data/met/crime-data-dashboard) using monthly counts for each crime type. In this range, there were 609 homicides, 369,963 violence with injury and 121,441 robbery of personal property crime counts. We acknowledge that a significant amount of crime counts will not be gang-related; however, they do reflect the overall landscape of violent crime in London.

*Out-of-vocabulary translation*

A limitation of previous work on drill music was the relatively high percentage of OOV terms. If too large a number of words or phrases are "unknown", the conclusions to be drawn from linguistic analyses are limited. Compared to a reference corpus of the most frequent 10k English words, almost one in five of the terms (18.64%) used in drill lyrics was not captured in previous investigations on this topic (Kleinberg and McFarlane, 2019). In order to mitigate that shortcoming, we retrieved the 300 unmatched words with the highest term frequency (e.g., *skeng*, *shank*, *splashed*) and manually translated these to standard English. Both authors independently used external resources (Tim Woods, no date; r/ukdrill, 2018; Tony Thorne, 2018) to look up translations of drill-specific terms (see Table 1). The translations of both authors were harmonized to one final translation (e.g., skeng → gun/weapon → weapon, shank → knife/weapon → knife, splashed → stabbed). For the subsequent analyses, we used the lyrics with the translated terms as replacement of the previous OOV terms.

Table 1. *Selection of OOV terms and translation*

| Term | Final translation |
|---|---|
| ting | gun (or: thing) |
| opps | enemies (members of another gang) |
| splash | stab |
| gyal | girl |
| cah | because |
| feds | cops, the police |
| whip | expensive car |
| ching | stab |
| peng | attractive girl |
| shank | knife |
| dando | abandoned property |

**METHOD**

*Linguistic trajectories*

We used the translation-corrected lyrics and extracted the sentiment trajectory using methods introduced in detail in previous work (Gao et al., 2016; Kleinberg et al., 2018; Soldner et al., 2019). Specifically, we weighted the sentiment in the presence of valence shifters (negators, amplifiers, de-amplifiers, adversative conjunctions) and iterated over the non-punctuated lyrics data using a moving window approach. The resulting vector consisted of the modified sentiment values and zero values (i.e. for words that do not hold a sentiment value) and was transformed to a standardized length to allow for comparison between trajectories. For the transformation, we used a discrete-cosine transform function with a low pass filter size of 10 to allow for a higher granularity in the trajectories. The trajectory method was used to examine the temporal development of sentiment use both on the corpus-level and the artist-level.

---

[3] The crime type categories for homicide, personal robbery and violence with injury are classified by the Home Office counting rules and classification standards.

*Corpus-level analysis*

The whole corpus of drill lyrics was time-ordered and concatenated to one long string, which was then used as input for the trajectory extraction method. To provide a smooth and interpretable overview of sentiment used in the whole corpus, we standardized the trajectory vector length to 10,000 (i.e. in steps of 0.01% of temporal progression). Since we were interested in the relative sentiment (i.e. the development of sentiment between a relative minimum and maximum) as well as the absolute sentiment, we also scaled the values to a range from -1.00 to +1.00. The highest sentiment for each trajectory is assigned a value of +1.00 and the lowest sentiment a value of -1.00. All intermediate values are scaled accordingly.

One aim of this investigation was to assess whether changes in positivity or negativity of the drill lyrics correspond to observed crime trends in London. The crime data (homicide, personal robbery, and violence with injury counts) were dated to the 10,000 bin trajectory vector, smoothed (using local regression with a span size of 0.30) for visual display and then scaled to a range of -1.00 to +1.00. That procedure allowed us to compare the trajectories visually. Note that quantitative comparisons are difficult since the vector lengths differ.

*Artist-level analysis*

For each artist, we extracted a trajectory of lyrics progression as follows: first, we order the lyrics by the time of appearance of the corresponding video on YouTube, starting with the oldest one. Second, we built a single long string for each artist consisting of all time-ordered lyrics and applied the sentiment trajectory extraction method. Third, all lyrics progression trajectories were standardized to a length of 1000 bins interpretable as steps of 0.1% of progression (i.e. the value at index 420 corresponds to 42% of lyrics progression of the artist).

Aside from understanding the trajectories, we were interested in assessing whether these shapes are distinct for each artist or whether some artists undergo a similar development in their lyrics. We, therefore, calculated the cosine similarity between the length-standardized trajectory vectors between the artists. The cosine similarity expresses the angle between two vectors and ranges from -1.00 (exactly opposite vectors) to +1.00 (exactly similar vectors). Cosine similarity of zero can be interpreted as the two vectors being independent. For the sake of interpretability, we label a cosine similarity smaller (larger) than -0.40 (+0.40) as dissimilar (similar), and all values between -0.40 and +0.40 as independent.

**RESULTS**

*Out-of-vocabulary reduction*

As expected, the manual translation of OOV terms into standard English resulted in a reduction of the percentage of unknown words. The OOV percentage was 13.08% (before translation: 18.64%) and 6.58% (before translation: 7.38%) using the 10k and 300k most frequent English words as a reference, respectively.

*Corpus-level analysis*

The trajectory of the whole London drill genre (Figure 1) reveals that as a whole, the drill lyrics contain predominantly negative sentiment. The sentiment never exceeds a neutral absolute sentiment threshold of zero. In relative terms, the lowest sentiment was present at the very start of the corpus (here: Dec. 2013), and we observe three local maxima at 14.67% (1[st] of Aug. 2016), 34.33% (28[th] of Feb. 2017) and 79.55% (9[th] of June 2018).

These findings suggest that over time, drill lyrics did not become more negative as a whole, albeit that in general drill music is characterized by negative sentiment words. Two decreasing trends are apparent in the trajectory data: between the two prominent peaks (34.33% and 79.55%), we see a longer downward trend to ever more negative sentiment up until 64.39% (28[th] of Jan 2018). The second downward trend follows the global maximum (79.55%) up until 93.58% (13[th] of Sept. 2018). For the remaining sentiment progression, we see an almost steady line.

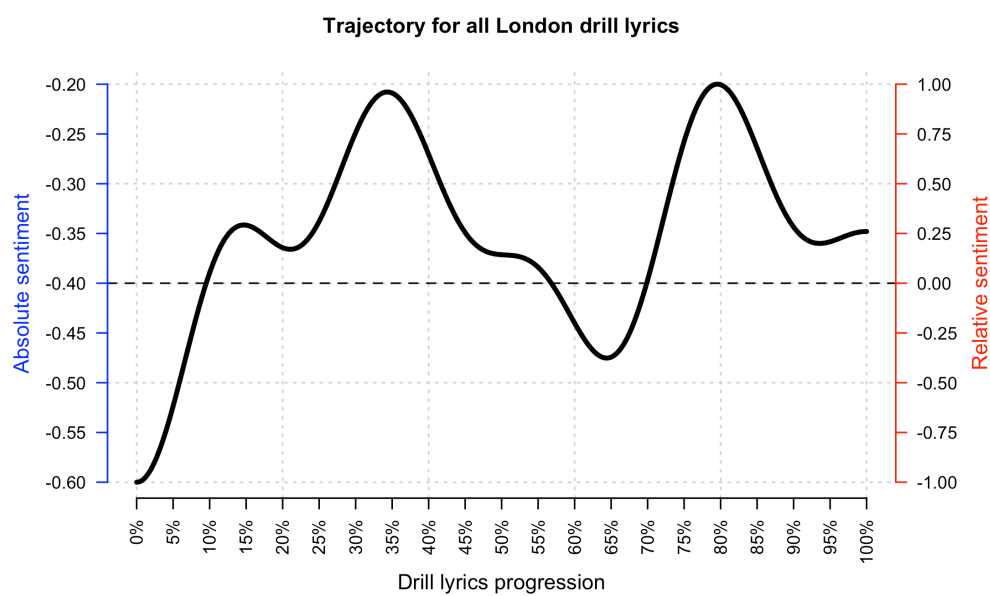

*Figure 1.* Absolute and relative temporal sentiment trajectory of the whole London drill corpus.

*Sentiment and crime trajectories*

The sentiment trajectory is compared to the crime count trajectories for London in Figure 2. For the crime count data, we see an overall increasing trend between the start and end of the date range (Dec. 2013 to Nov. 2018). Robbery from the person increases consistently from shortly after the start of the date range to peak at 55.88% (Nov. 2017) on the temporal dimension. Violence with injury counts peak at 45.69% (July 2017) but fluctuate in waves with another peak at 86.14% (July 2017). Homicides have two noteworthy peaks at 48.62% (Aug. 2017) and 79.42% (Mar. 2018).

If negative sentiment in drill music were related to increases in either crime type, we would expect anti-cyclic patterns: relatively negative lyrics would then correspond to or be followed by relatively higher crime counts. Since any suchlike relationship would require a lag in time, we find only one potential relationship. The local minimum in sentiment at 64.39% (28[th] of Jan 2018) is followed by a peak in homicide at 79.42% (Mar. 2018).

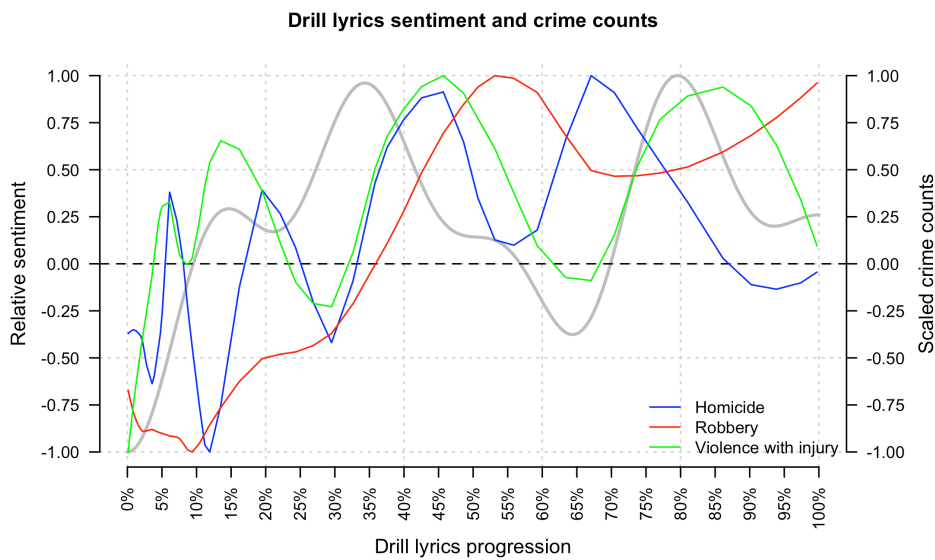

*Figure 2.* Relative temporal sentiment trajectory (grey) and scaled crime counts.

*Artist-level trajectories*

The trajectories for six artists with the most songs in our corpus (here at least 23 songs) are shown in Figure 3. We were interested in whether the temporal developments of these artists were distinct patterns (i.e. each artist's trajectory was unique) or displayed similarities with other artists. The cosine similarities between the artists' temporal trajectory vectors (Table 2) show that three artists share similar trajectories:

- "67" and "Harlem Spartans" both show an increase-decrease pattern at 10-15% and 25-30% progression, respectively. Both stabilize in sentiment between 40-70% and then decrease until ca. 80%, after which they increase towards the end.
- "410" and "Abra Cadabra" show their highest sentiment at the very start of their temporal trajectory and display a pattern of four peaks. Both experience their lowest sentiment at ca. 40-45%.
- "Abra Cadabra" and "Zone 2" share an elongated decrease (with a small bump for "Abra Cadabra" at ca. 25%) after a start with their highest sentiment up until ca. 45%.

Nine of the 15 comparisons resulted in independent trajectories, suggesting that they are neither similar nor dissimilar. For "Harlem Spartans" and "Abra Cadabra" we observed divergent trajectories suggesting that their progression was close to opposite to that of the other.

Table 2. *Cosine similarities between the trajectory vectors of the top 6 artists.*

|  | "67" | "410" | "Harlem Spartans" | "Abra Cadabra" | "Zone 2" | "Headie One" |
|---|---|---|---|---|---|---|
| "67" | 1.00 | independent | similar | dissimilar | independent | dissimilar |
| "410" | -0.05 | 1.00 | independent | similar | independent | independent |
| "Harlem Spartans" | **0.68** | -0.10 | 1.00 | dissimilar | independent | independent |
| "Abra Cadabra" | *-0.32* | **0.44** | *-0.60* | 1.00 | similar | independent |
| "Zone 2" | 0.21 | 0.04 | -0.01 | **0.55** | 1.00 | independent |
| "Headie One" | *-0.37* | 0.12 | -0.07 | -0.05 | -0.11 | 1.00 |

*Note.* Lower diagonal: read as the cosine similarity between [row] and [column] was X. The upper diagonal offers an interpretation of the values. Cosine similarity values interpreted as "similar" are in **bold**; values that are interpreted as dissimilar are in *italics*.

**DISCUSSION**

In this paper, we intended to expand the study of drill lyrics as a phenomenon of a highly localized, niche sub-culture with three key contributions. (1) We utilized a trajectory method to understand the development of language use for the whole genre and (2) compared these findings to actual crime data. (3) We also assessed the development of individual artists and tested whether these are distinct patterns or whether some artists develop similarly.

*Main findings*

This investigation was the first to assess the development of the whole London drill music sub-genre. An analysis of the development of sentiment use in all lyrics reveals that as a whole, drill music is predominantly negative in sentiment. The trajectory of the genre reveals that it becomes more positive over time but still remains negative in general. With two peaks at ca. 35% and 80% of the corpus time, the current state of drill music in the UK is less positive than at its peak but still more positive than it was in late 2013 and early 2014.

An initial look at the relationship between crime and sentiment did not support the idea of the negativity of drill lyrics as a leading indicator (or even incitement) for real-life serious crimes. Only one candidate for such a relationship was found with a peak in homicides around March 2018 following a low-point in sentiment in late January 2018. It is important to note that causality in any relationship of observational data is practically impossible to ascertain. As such, and absent any further data and analysis, we cannot interpret these results as support for the hypothesis that sentiment can be used as an indicator of real-life crime.

At the artist-level, our work suggests that some drill artists have a similar developmental trajectory. While more data are needed to consolidate these findings, the initial results provide a basis for future work. The question of whether trajectories are unique or not has been examined in previous research on novels (Jockers, 2015; Gao et al., 2016), vlogs (Kleinberg et al., 2018), TED talks (Tanveer et al., 2018) and partisan news coverage (Soldner et al., 2019). However, these investigations looked at the individual document-level rather than style as a construct of the individuals producing the text. Future work could expand on the current approach and assess whether style as a dynamic construct modelled through trajectories is consistent among individuals or not.

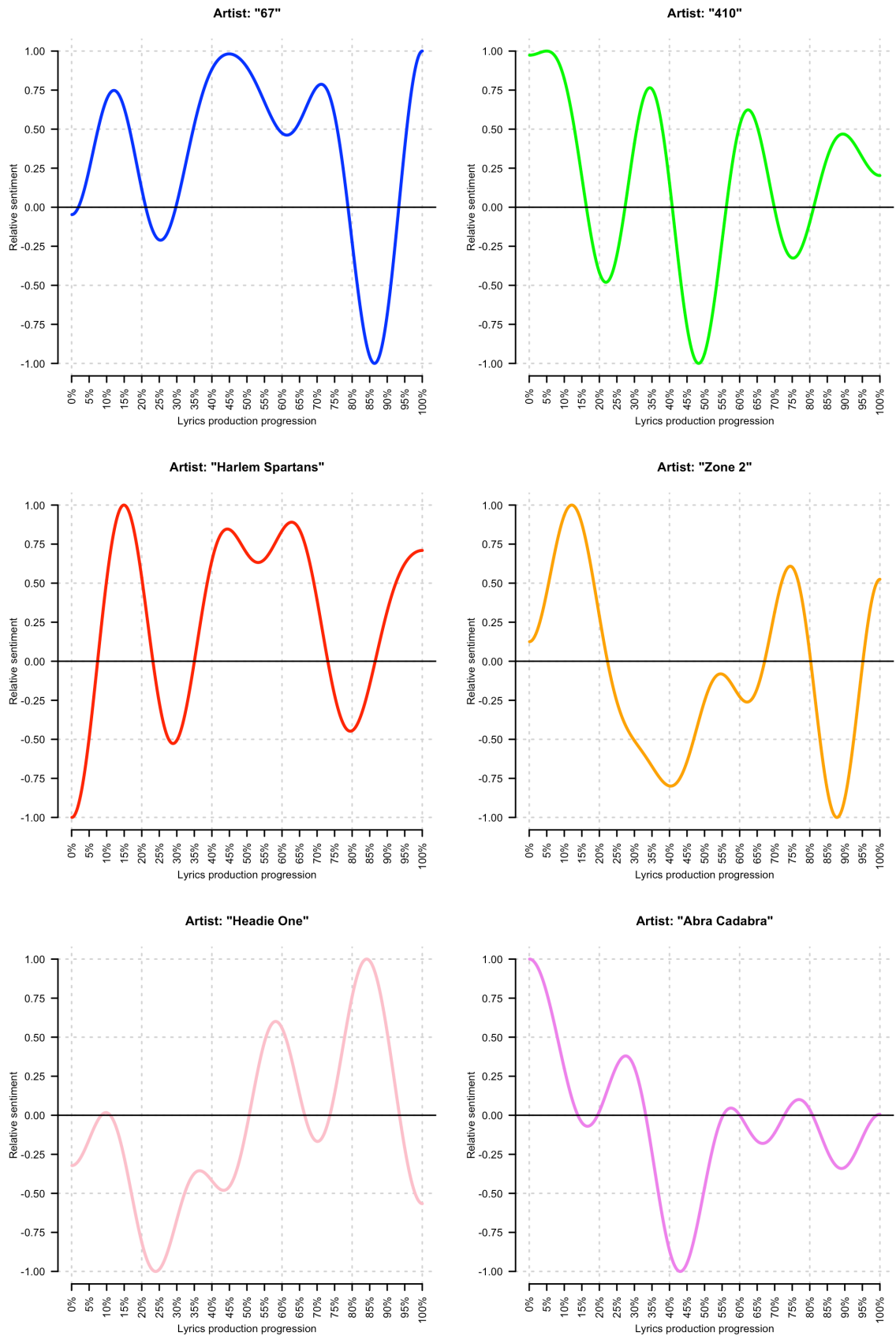

*Figure 3.* Relative temporal sentiment trajectories of the artists' progression.

*Ideas for future work*

The current work further adds to an understanding of drill music and its potential relationship to crime in London. However, the following limitations merit attention and could be addressed in future research. First, the manual translation approach of the current study has reduced the percentage of unknown words but did not address the issue of the highly-contextualized meaning of some terms. Although known to a standard English lexicon, many words hold a different meaning in the drill-music context compared to everyday English. For example, the word "ting" has various meanings, such as "girl', "gun" or "thing". Future work might want to look at a more comprehensive translation approach using not only out-of-vocabulary terms but also those that carry a different meaning in the drill context.

Second, an underlying premise of this investigation is that online behaviour (here: posting videos, interacting with videos) has a relationship with offline events or behaviour. Although practically all examinations of online behavioural data make that assumption, it is far from established or supported by research. Future work in the area of computational social science and behavioural data science could focus on these aspects to provide a solid base upon which the field, and investigations such as the current one, can build.

Third, drill music is not a phenomenon unique to the UK. Recently, the genre has gained attention in the Netherlands using the same (alleged) drill-violence nexus as in the UK (Vugts and van Dun, 2019). Recent stabbing incidents have been connected to drill music (de Haan, 2020) while research finding such a connection is still lacking. With future work on the matter across countries, government and local policy-makers can start devising evidence-based intervention strategies.

*Questions ahead and the bigger picture*

Given the scarcity of empirical work on the drill-violence nexus, it is challenging to determine whether current government policies – be they restrictive on drill music as in the UK (Dearden, 2018; Waterson, 2018) or more tentative as put forward by the major of Amsterdam (Koops, 2019) - are supported by evidence or not. However, given the urgency of the problem, two broader questions merit attention for the research agenda.

First, if – and this needs further corroboration beyond the two pieces published so far – drill music on YouTube does not function as an accelerator of violence and is not used by the artists to incite or provoke violence, then a potential explanation for the extensive use of the video-sharing platform could be rather mundane. YouTube will remunerate drill artists for the number of views their videos receive, some of which have millions of views. Obviously, developing and uploading music content that receives more views, comments and likes increases the profile and popularity of the artists and also has the potential to generate significant revenue. It would be interesting to test whether financial rather than gang-related motives are the main driver for sharing and producing high-quality videos.

Second, our current work fails to find support for the overarching narrative that drill music is related to violence. But it is essential to consider that we used a computational approach throughout (except for some manual term translation). While such an approach is ultimately more useful to stakeholders who make decisions quickly and need to extract information from vast amounts of data, the computational approach simplifies nuances. It is possible that there are blind spots in the drill music sphere that we do not capture. These include aggression displayed in visual cues, code language that differs in meaning from standard English, wider interaction with content (e.g. comments on YouTube), or the use of other platforms (e.g. Instagram). The problem is too significant not be given attention, and we encourage others to challenge our findings and look for blind spots that we might have missed.

**CONCLUSION**

In this paper, we revisited the question of whether there is evidence to support the conjecture that drill music and gang violence are inextricably linked. Moreover, we argue that current policy is somewhat disconnected from the evidence base, and more reasonable assumptions can now be used to revise policy towards drill music. We have provided some evidence to suggest drill music lyrics have not become more negative but, in fact, have become more positive over time. Individual drill artists have similar sentiment trajectories to other artists in the drill genre, which might help in understanding the problem further. Finally, our work did not find evidence of a relationship between drill music and 'real-life' violence when compared to three violent crime trajectories in London. Our findings can be used to continue to build the evidence base that informs the police and social media policymakers regarding the lyrical content of drill music videos. While the existing consensus in the UK is that drill music videos should be removed from social media platforms, we argue that further consideration should be given to this potentially damaging approach.

We do not – and cannot - claim that drill music videos and their unique content have not been contributory to some instances of gang-related violence. Similarly, we also acknowledge that the police may have contextual information or intelligence not available to the public (and researchers) that supports some linkages between drill music artists and gang violence. Nevertheless, at this point in time, to the best of our knowledge, there is no publicly available empirical evidence that drill music language incites violent crime.